\documentclass[prl,twocolumn,superscriptaddress,amsmath,amssymb]{revtex4}
\usepackage{graphicx}% Include figure files
\usepackage{dcolumn}% Align table columns on decimal point
\usepackage{bm}% bold math
\usepackage{color}% perhaps for xfig exports needed
\usepackage{epsfig}%perhaps necessary for xfig exports
\usepackage{amssymb}
\usepackage{amsmath}

\begin{document}

\newcommand{\commute}[2]{\left[#1,#2\right]}
\newcommand{\bra}[1]{\left\langle #1\right|}
\newcommand{\ket}[1]{\left|#1\right\rangle }
\newcommand{\anticommute}[2]{\left\{  #1,#2\right\}  }
\renewcommand{\arraystretch}{2}

\title{Spin-echo of a single electron spin in a quantum dot}

\author{F.H.L. Koppens}
\affiliation{Kavli Institute of NanoScience Delft, P.O. Box 5046, 2600 GA Delft, The Netherlands}
\author{K.C. Nowack}
\affiliation{Kavli Institute of NanoScience Delft, P.O. Box 5046, 2600 GA Delft, The Netherlands}
\author{L.M.K. Vandersypen}
\affiliation{Kavli Institute of NanoScience Delft, P.O. Box 5046, 2600 GA Delft, The Netherlands}

\begin{abstract}
We report a measurement of the spin-echo decay of a single electron spin confined in a semiconductor quantum dot. When we tip the spin in the transverse plane via a magnetic field burst, it dephases in 37 ns due to the Larmor precession around a random effective field from the nuclear spins in the host material. We reverse this dephasing to a large extent via a spin-echo pulse, and find a spin-echo decay time of about 0.5 $\mu$s at 70 mT. These results are in the range of theoretical predictions of the electron spin coherence time governed by the dynamics of the electron-nuclear system.
\end{abstract}

\maketitle 

%Introduction

%Isolated electron spins in a semiconductor are natural two-level quantum systems which can have very long coherence times. This permits studies of their fundamental quantum mechanical behavior, and also holds promise for quantum information processing applications. 
%For instance, decoherence times of hundreds of microseconds have been obtained for ensembles of electron spins bound to phosphorous donors in silicon\cite{gordon58}, where a spin-echo technique was used in order to remove systematic dephasing due to inhomogeneous broadening.
%Rather long coherence times are also expected for spins confined in gate-defined quantum dots, which is the system we really want to use, as it has proven success in GaAs. In all III-V systems, however, the slow intrinsic decoherence is obscured by a much faster systematic dephasing because the electron spin experiences a random, unknown nuclear field. If want to use this system for qubit, need to solve this. because nuclei evolve slowly, can be solved. either measure or set the nuclear field, or use echo. has been implemented for two spins, by using exchange interaction.

Isolated electron spins in a semiconductor can have very long coherence times, which permits studies of their fundamental quantum mechanical behavior, and holds promise for quantum information processing applications \cite{lossdivincenzo}. For ensembles of isolated spins, however, the slow intrinsic decoherence is usually obscured by a much faster systematic dephasing due to inhomogeneous broadening \cite{dutt05,bracker05a}. The actual coherence time must then be estimated using a spin echo pulse that reverses the fast dephasing\cite{hahn56}. In this way, decoherence times as long as hundreds of microseconds have been measured for ensembles of electron spins bound to phosphorous donors in silicon \cite{gordon58}.

For a single isolated spin there is no inhomogeneous broadening due to averaging over a spatial ensemble. Instead, temporal averaging is needed in order to collect sufficient statistics to characterize the spin dynamics. In some cases, this averaging can also lead to fast apparent dephasing that can be (largely) reversed using a spin-echo technique. This is possible when the dominant influence on the electron spin coherence fluctuates slowly compared to the electron spin dynamics, but fast compared to the required averaging time. Such a situation  is common for an electron spin in a GaAs quantum dot where the hyperfine interaction with the nuclear spins gives rise to an effective slowly fluctuating nuclear field, resulting in a dephasing time of about tens of nanoseconds \cite{khaetskii02,merkulov02,johnson05a,koppens05,petta05}. The effect of the low-frequency components of the nuclear field can be reversed to a large extent by a spin-echo technique. For two-electron spin states, this was demonstrated by rapid control over the exchange interaction between the spins \cite{petta05}. The application of a spin-echo technique on a single electron spin is required when using the spin in a GaAs quantum dot as a qubit. Furthermore, erasing fast dephasing allows for a more detailed study on the remaining decoherence processes such as the rich dynamics of the electron-nuclear system \cite{sousa03a,coish04,yao06,coish07c}. 

%
% Confining single electron spins in gate-defined quantum dots offers the advantage of scalable device fabrication and in-situ tunability of the coherent coupling between the two spins is readily available \cite{elzerman03,petta05}.
%
%Also in this system, the dephasing dynamics can be reversed to a large extent by a spin-echo technique, which was demonstrated for two-electron spin states by fast of the exchange interaction. Operating a single electron spin as a qubit requires the applyication of a spin-echo technique on a single spin as well, but it also allows for a more detailed study of the rich dynamics of the remaining decoherence processes. 
% Long coherence times are expected for spins in these structures as well \cite{khaetskii00,golovach04}, but for all III-V semiconductors, the hyperfine interaction with the nuclear spins forms an important dephasing source \cite{khaetskii02,merkulov02,johnson05a,koppens05}. Still, the dephasing dynamics which can be seen as a Larmor precession of the spin around an unknown static effective nuclear field, can be reversed to a large extent by a spin-echo technique. This is required when operating a single spin as a qubit, but also allows for a more detailed study of the rich dynamics of the remaining nuclear-induced decoherence processes. 
%
%is the subject of our study.  Confining single electron spins in gate-defined quantum dots offers the advantage of scalable device fabrication and in-situ tunability of the coherent coupling between the two spins is readily available \cite{elzerman03,petta05}
%
%ADD MENTION OTHER Q SYSTEMS

\begin{figure}[b]
%\begin{figure}[p]
          \includegraphics[scale=1]{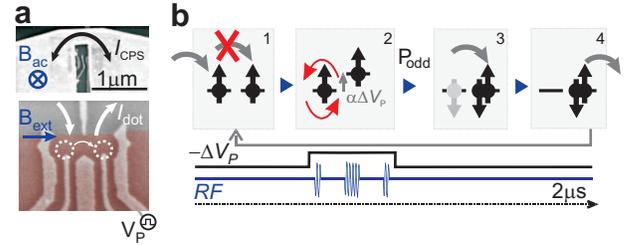}
\caption{ 
(color online) a) Bottom: Scanning electron microscope (SEM) image of the Ti/Au gates on top of a GaAs/AlGaAs heterostructure containing a two-dimensional electron gas 90 nm below the surface. White arrows indicate current flow through the two coupled dots (dotted circles). The gate labeled with $V_p$ is connected to a homemade bias-tee (rise time 150 ps) to allow fast pulsing of the dot levels. 
Top: SEM image of the on-chip coplanar stripline, separated from the surface gates by a 100-nm-thick dielectric. Due to the geometry of the stripline, the oscillating field with amplitude $B_\mathrm{ac}$ and frequency $f_\mathrm{ac}$ is generated primarily perpendicular to the static field $B_\mathrm{ext}$, which is applied in the plane of the two-dimensional electron gas. 
b) Schematic of the electron cycle (time axis not on scale). The voltage $\Delta V_p$ (with lever arm $\alpha$) on the gate detunes the dot levels during the manipulation stage (applied bias is 1.5 mV).}
\label{oscillations}
\end{figure}

Here, we report the use of a spin-echo technique for probing the coherence of a single electron spin confined in an electrostatically defined GaAs quantum dot (shown in Fig. \ref{oscillations}a). The spin is manipulated using electron spin resonance, as reported in \cite{koppens06}. We first realize a two-pulse experiment akin to Ramsey interference, whereby fringes develop when the relative phase between the pulses is varied. By varying the delay time between the pulses, we measure a dephasing time $T_2^*$ of about 37 ns. When using a spin-echo technique, a spin-echo decay time $T_\mathrm{2,echo}$ is obtained of about 0.5 $\mu$s. This is more than a factor of ten longer than the Ramsey decay time, indicating that the echo pulse reverses the dephasing to a large extent. These findings are consistent with theoretical predictions for this system \cite{khaetskii02,merkulov02,sousa03a,coish04,yao06,deng06}, as well as with earlier echo measurements on two-electron spin states in a similar quantum dot system \cite{petta05}, and with mode locking measurements of single spins in an ensemble of self-assembled quantum dots \cite{greilich06b}.

\begin{figure}[rt]
%\begin{figure}[p]
          \includegraphics[scale=1]{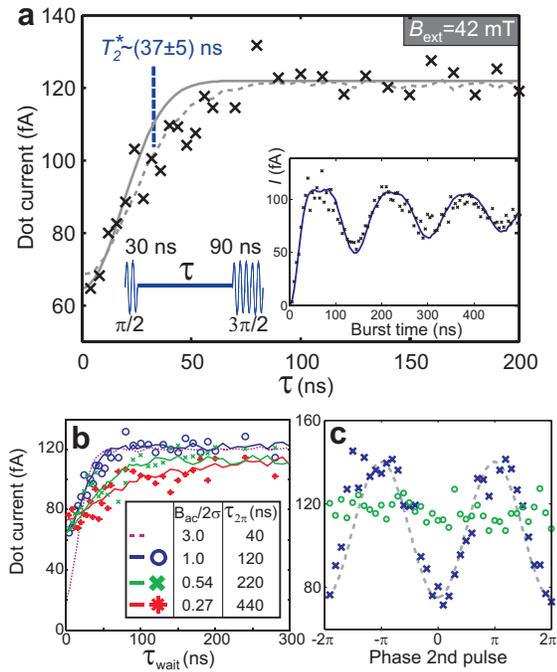}
\caption{(color online) a)  Ramsey signal as a function of free-evolution time $\tau$. Each data point reflects a current measurement averaged over 20 seconds at constant $B_\mathrm{ext}=$42 mT, $f_\mathrm{ac}=210$ MHz, $B_\mathrm{ac}=3$ mT (as shown in the inset, this gives a Rabi period $\tau_\mathrm{2\pi}$ of 120 ns; see \cite{koppens06} for further details). In order to optimize the visibility of the decay, the second pulse is a $3\pi/2$-pulse instead of the usual $\pi/2$-pulse (the measured signal depends on $P_\mathrm{odd}$, see main text). 
Solid line: Gaussian decay with $T_2^*=30$ ns, corresponding to $\sigma=1.5$ mT.
Dotted line: numerically calculated current. First $P_\mathrm{odd}$ is computed taking $\sigma=1.5$ mT, and then the current is derived as $I_\mathrm{dot} = P_\mathrm{odd}(m+1)80+23$ fA ($m$ and offset due to background current obtained from fit). A current of 80 fA corresponds to one electron transition per 2 $\mu s$ cycle, and $m$ is the additional number of electrons that tunnels through the dot on average before the current is blocked again. Here, we find $m=1.44$; the deviation from the expected $m=1$ is not understood and discussed in \cite{koppens06}. 
b) Measured and numerically calculated Ramsey signal for a wide range of driving fields. The simulations assume $\sigma=1.5$ mT, and estimate the current as $P_\mathrm{odd}(m+1)80+23$ fA ($m=1.5$) for $\tau_{2\pi}$=40-220 ns, and as $P_\mathrm{odd}(m+1)80+43$ fA ($m=1.5$) for $\tau_{2\pi}$=440 ns.  
c) Ramsey signal as a function of the relative phase between the two RF bursts for $\tau=$ 10 (crosses) and 150 ns (circles). Gray dashed line is a best fit of a cosine to the data.}
     
\end{figure}

The measurement scheme is depicted in Fig.\,1a. Two quantum dots are tuned such that one electron always resides in the right dot and a second electron can flow through the two quantum dots only if the spins are anti-parallel. For parallel spins, the second electron cannot enter the right dot due to the Pauli exclusion principle, and is blocked in the left dot \cite{remark_blockade}. This allows us to initialize the system in a mixed state of $\ket{\uparrow\uparrow}$ and $\ket{\downarrow\downarrow}$ (stage 1), although from now on, we assume the initial state is $\ket{\uparrow\uparrow}$, without loss of generality. Next the electron spins are manipulated with a sequence of RF bursts \cite{vectorsource} (stage 2), while a voltage pulse $\Delta V_\mathrm{p}$ is applied to one of the gates so that tunneling is prohibited regardless of the spin states. Once the pulse is removed, electron tunneling is allowed again, but only for anti-parallel spins (stages 3 and 4). The entire cycle lasts 2 $\mu$s and is continuously repeated, resulting in a current flow which reflects the average probability $P_\mathrm{odd}$ to find anti-parallel spins at the end of stage 2. 

We first use this scheme to measure the dephasing of the spin via a Ramsey-style experiment (see inset Fig.\,2a for the RF pulse sequence). A $\pi/2$-pulse is applied to create a coherent superposition between $\ket{\uparrow}$ and $\ket{\downarrow}$, after which the spin is allowed to freely evolve for a delay time $\tau$ (for now, we reason just in a single-spin picture \cite{onespin}). Subsequently, a $3\pi/2$-pulse is applied, with a variable phase. Ideally, if both RF pulses have the same phase, the spin is rotated back to $\ket{\uparrow}$, and the system returns to spin blockade. If the phases of the two pulses are $180^\circ$ out of phase, the spin is instead rotated to $\ket{\downarrow}$, and the blockade is lifted. Fig.\,2c shows that for small $\tau$, the probability to find $\ket{\downarrow}$ indeed oscillates sinusoidally as a function of the relative phase between the two RF pulses. This is analogous to the well-known Ramsey interference fringes. For large $\tau$, however, the spin completely dephases during the delay time, and the fringes disappear (Fig.\,2c). We see that the maximum contrast for the effect of dephasing is obtained for two pulses with the same phase. The signal for this case is shown in Fig.\,2a, as a function of $\tau$. We find that the signal saturates on a timescale of $T_2^*\sim$ 37 ns (obtained from a Gaussian fit, see below), which gives a measure of the dephasing time. 

The observed Ramsey decay time is the result of the hyperfine interaction between the electron spin and the randomly oriented nuclear spins in the host material. The coupling Hamiltonian is given by $H_\mathrm{hf}={\bf S} \cdot {\bf h}= S_z h_z+\frac{1}{2}(S_+ h_- + S_- h_+)$ where ${\bf S}$ is the spin-1/2 operator for the electron spin and ${\bf h}=\sum_i A_i \mathrm{\bf I_i}$, with ${\bf I}_i$ the operator for nuclear spin $i$ and $A_i$ the corresponding hyperfine coupling constant. The $S_\mathrm{z} h_\mathrm{z}$ term in the Hamiltonian can be seen as a nuclear field $B_{N,z}$ in the z-direction that modifies the Larmor precession frequency of the electron spin (the other two terms are discussed below in the context of spin-echo). This nuclear field fluctuates in time with a Gaussian distribution with width $\sigma$ and a typical correlation time $\sim100 \ \mu s-1\ s$. This is much longer than the 2 $\mu s$ cycle time, but much shorter than the averaging time for each measurement point ($\sim$ 20 seconds). Averaging the precession about $B_{N,z}$ during time $\tau$ over a Gaussian distribution of nuclear fields, gives a Gaussian coherence decay $\int^{\infty}_{-\infty} \frac{1}{\sqrt{2\pi}\sigma} e^{-B_\mathrm{N,z}^2/2\sigma^2} \cos(g\mu_b B_\mathrm{N,z}\tau/\hbar) dB_{N,z} = e^{-(\tau/T^*_2)^2}$, with $T^*_2=\sqrt{2}\hbar/g\mu_b\sigma \sim 30$ ns \cite{khaetskii02,merkulov02} (assuming $\sigma=$1.5 mT, extracted from the Rabi oscillations, see \cite{koppens07}). This decay is plotted in Fig. 2a (solid line). However, the observed Ramsey signal cannot be compared directly with this curve because we have to take into account the effect of the nuclear field during the $\pi/2$ and $3\pi/2$-pulses as well. Essentially, $B_\mathrm{N,z}$ shifts the electron spin resonance condition and as a result, the fixed-frequency RF pulses will be somewhat off-resonance which decreases the fidelity of the rotations.

We include these effects in a simulation of the spin dynamics, and consider from here on not just a single spin but the actual two-spin system. At the end of the cycle, the two-spin state is given by $\psi(\tau,B_\mathrm{L,R}) = U_{\frac{3\pi}{2}}^L(B_\mathrm{L})U_{\frac{3\pi}{2}}^R(B_\mathrm{R})V_\tau^\mathrm{L}
 (B_\mathrm{L})V_\tau^\mathrm{R}(B_\mathrm{R})U_{\frac{\pi}{2}}^L(B_\mathrm{L})U_{\frac{\pi}{2}}^R(B_\mathrm{R})\ket{\uparrow\uparrow}$. 
Here, $U_{\theta}^{L,R}(B_\mathrm{L,R})$ is the single spin time-evolution operator (for an intented $\theta$-rotation) resulting from the driving field and the $z$-component of the nuclear fields in the left and right dot, $B_\mathrm{L}$ and $B_\mathrm{R}$. The operator $V_\tau^\mathrm{L,R}(B_\mathrm{L,R})$ represents the single spin evolution during a time $\tau$ in the presence of the nuclear field only. We can then compute $P_\mathrm{odd}$ at the end of the pulse sequence, averaging over two independent Gaussian distributions of nuclear fields in the left and right dot:
\begin{eqnarray*}
P_\mathrm{odd}(\tau)= \frac{1}{2\pi\sigma^2} \int\!\!\int e^{-(\frac{B_L^2+B_R^2}{2\sigma^2})} \widetilde{P}_\mathrm{odd}(\tau,B_{L,R}) \; dB_LdB_R\;;\\
\widetilde{P}_\mathrm{odd}(\tau,B_\mathrm{L,R})=\left|\bra{\psi(\tau,B_\mathrm{L,R})}{\uparrow\downarrow}\rangle \right|^2+\left|\bra{\psi(\tau,B_\mathrm{L,R})}{\downarrow\uparrow}\rangle\right|^2.
\end{eqnarray*}
This numerically calculated $P_\mathrm{odd}(\tau)$ is plotted in Fig.\,2a (dotted line), after rescaling in order to convert $P_\mathrm{odd}$ to a current flow $I_\mathrm{dot}$ (see caption). We see that the predicted decay time is longer when the rotations are imperfect due to resonance offsets. This is more clearly visible in Fig.\,2b, where the computed curves are shown together with Ramsey measurements for a wide range of driving fields. The experimentally observed Ramsey decay time is longer for smaller $B_\mathrm{ac}$, in good agreement with the numerical result. This effect can be understood by considering that a burst doesn't (much) rotate a spin when the nuclear field pushes the resonance condition outside the Lorentzian lineshape of the excitation with width $B_\mathrm{ac}$. If the spin is not rotated into a superposition, it cannot dephase either. As a result, the cases when the nuclear field is larger than the excitation linewidth do not contribute to the measured coherence decay. The recorded dephasing time is thus artificially extended when long, low-power RF bursts are used ($B_\mathrm{ac}/2\sigma \lesssim 1$). However, in Fig.\,2a, this is only a small effect.

\begin{figure}[tr]
%\begin{figure}[p]
          \includegraphics[scale=1]{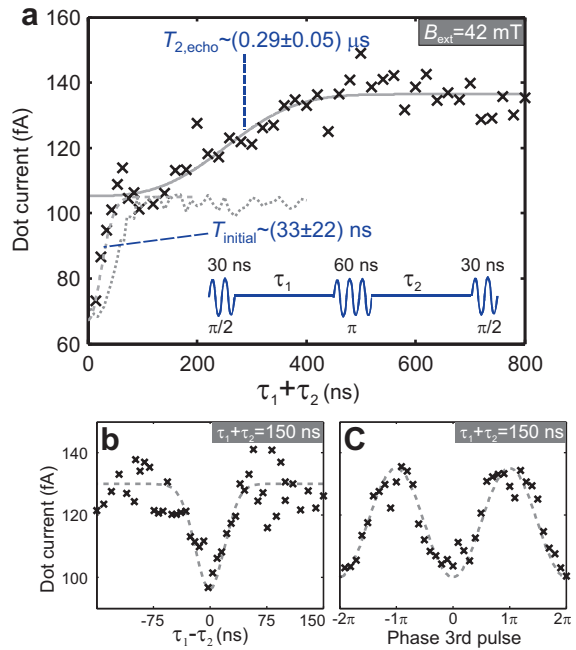}
     \caption{(color online) a) Spin-echo signal as a function of total free-evolution time $\tau_1+\tau_2$. Each data point represents the current through the dots averaged over 20 seconds at constant $B_\mathrm{ext}=42$ mT, $f_\mathrm{ac}=210$ MHz, $B_\mathrm{ac}=3$ mT. Dashed line: best fit of a Gaussian curve to the data in the range $\tau_1+\tau_2=0-100$ ns.
Solid line: best fit of $e^{-((\tau_1+\tau_2)/T_\mathrm{2,echo})^3}$ to the data in the range $\tau_1+\tau_2=100-800$ ns. Dotted line: numerically calculated dot current $P_\mathrm{odd}(m+1)80+25$ fA, taking $\sigma=1.5$ mT in both dots and $m=1.83$. Considerable scattering of the data points is not due to the noise of the measurement electronics (noise floor about 5 fA), but caused by the slow evolution of the statistical nuclear field. Inset: spin-echo pulse sequence.  b) Spin-echo signal as a function of $\tau_1-\tau_2$. Dashed line: best fit of a Gaussian curve to the data. c) Spin-echo signal for $\tau_1+\tau_2=150$ ns as a function of the relative phase between the first two and third pulse. Dashed line is the best fit of a cosine to the data. }          
\end{figure}

We now test to what extent the electron spin dephasing is reversible using a spin-echo pulse.
The pulse sequence we apply is depicted in the inset of Fig.\,3a, and the measured signal as a function of the total free-evolution time $\tau_1+\tau_2$ is shown in the main panel of Fig.\,3a. We see immediately that the spin-echo decay time, $T_\mathrm{2,echo}$, is much longer than the dephasing time, $T_2^*$. 

This is also clear from the data in Fig. 3c, which is taken in a similar fashion as the Ramsey data in Fig. 2c, but now with an echo pulse applied halfway through the delay time. Whereas the fringes were fully suppressed for a 150 ns delay time without an echo pulse, they are still clearly visible after 150 ns if an echo pulse is used. As a further check, we measured the echo signal as a function of $\tau_1-\tau_2$ (Fig.\,3b). As expected, the echo is optimal for $\tau_1=\tau_2$ and deteriorates as $|\tau_1-\tau_2|$ is increased. The dip in the data at $\tau_1-\tau_2=0$ has a half width of $\sim$27 ns, similar to the observed $T_2^*$. 

Upon closer inspection, the spin-echo signal in Fig.\,3a reveals two types of decay.  First, there is an initial decay with a typical timescale of 33 ns (obtained from a Gaussian fit), which is comparable to the observed Ramsey decay time when using the same $B_\mathrm{ac}$. This fast initial decay occurs because the echo pulse itself is also affected by the nuclear field. As a result it fails to reverse the electron spin time evolution for part of the nuclear spin configurations, in which case the fast dephasing still occurs, similar as in the Ramsey decay. To confirm this, we calculate numerically the echo signal, including the effect of resonance offsets from the nuclear fields, similar as in the simulations of the Ramsey experiment. We find reasonable agreement of the data with the numerical curve (dotted line in Fig.\,3a), regarding both the decay time and the amplitude. 

The slower decay in Fig.\,3a corresponds to the loss of coherence that cannot be reversed by a perfect echo pulse. We extract the spin-echo coherence time $T_\mathrm{2,echo}$ from a best fit of $a+be^{-((\tau_1+\tau_2)/T_\mathrm{2,echo})^3}$ \cite{sousa03a,yao06} to the data ($a,b,T_\mathrm{2,echo}$ are fit parameters) and find $T_\mathrm{2,echo}=(290 \pm 50$) ns at $B_\mathrm{ext}$=42 mT (see Fig.\,3a, solid line).  We note that the precise functional form of the decay is hard to extract from the data, but fit functions of the form $a+be^{-(\tau/T)^d}$ with $d$ between 2 and 4 give similar decay times. 

Measurements at higher $B_\mathrm{ext}$ are shown in Fig.\,4a,b. Here, experiments were only possible by decreasing the driving field and as expected, we thus find a longer initial decay time, similar as seen in Fig.\,2b for Ramsey measurements. The longer decay time from which we extract $T_\mathrm{2,echo}$ tends to increase with field, up to 0.44 $\mu$s at $B_\mathrm{ext}$=70 mT. This is roughly in line with the spin echo decay time of $1.2\,\mu$s observed for two-electron spin states at $B_\mathrm{ext}$=100 mT \cite{petta05}. 

The field-dependent value for $T_\mathrm{2,echo}$ we find is more than a factor of 10 longer than $T_2^*$, which is made possible by the long correlation time of the nuclear spin bath. We now examine what mechanism limits $T_\mathrm{2,echo}$. The $z$-component of the nuclear field can change due to the spin-conserving flip-flop terms $H_\mathrm{ff} = \frac{1}{2}(S_+ h_- + S_- h_+)$ in the hyperfine Hamiltonian ${\bf S} \cdot {\bf h}$, and due to the dipole-dipole interaction between neighbouring nuclear spins. Direct electron-nuclear flip-flop processes governed by $H_\mathrm{ff}$ are negligible at the magnetic fields used in this experiment, because of the energy mismatch between the electron and nuclear spin Zeeman splitting. However, the energy-conserving higher-order contributions from $H_\mathrm{ff}$ can lead to flip-flop processes between two non-neighboring nuclear spins mediated by virtual flip-flops with the electron spin \cite{coish04,shenvi05,deng06,yao06}. It is predicted that this hyperfine-mediated nuclear spin dynamics can lead to a field dependent free-evolution decay of about 0.1-100 $\mu$s for the field range 1-10 T \cite{deng06,yao06,coish07c}. Interestingly, some theoretical studies \cite{shenvi05,yao06} have shown that this type of nuclear dynamics is reversible (at sufficiently high field) by an echo-pulse applied to the electron spin. The coherence decay time due to the second possible decoherence source, namely the dipole-dipole interaction, is expected to be 10-100 $\mu$s \cite{sousa03a}, independent of magnetic field (once $B_\mathrm{ext}$ is larger than $\sim 0.1$ mT, which is the dipole field of one nucleus seen by its neighbour). 

Also decoherence mechanisms other than the interaction with the nuclear spin bath must be considered. One possibility is spin-exchange with electrons in the reservoir via higher order tunneling processes. However, we  expect that the typical timescale of this process is very long because (during the manipulation stage) the energy required  for one of the electrons to be promoted to a reservoir ($>100\,\mu$eV) is much larger than the tunnel rate ($<0.1\,\mu$eV). In principle, the Heisenberg coupling $J$ between the electron spins in the two quantum dots could also lead to decoherence, but during the manipulation stage, we expect that $J$ is very small due to the large level detuning. Altogether, the most likely limitation to the observed $T_\mathrm{2,echo}$ is hyperfine-mediated flip-flops between any two nuclear spins. 

To conclude, we have performed time-resolved measurements of the dephasing of a single electron spin in a quantum dot caused by the interaction with a quasi-static nuclear spin bath. We have largely reversed this dephasing by the application of a spin-echo technique. The echo pulse extends the decay time of the electron spin coherence by more than a factor of ten. We obtain a $T_\mathrm{2,echo}$ of 0.29 $\mu$s and 0.44 $\mu$s at magnetic fields of 42 and 70 mT respectively. While even longer coherence times are expected at higher magnetic fields, the observed decay times are already sufficiently long for further exploration of electron spins in quantum dots as qubit systems. 

%Mechanisms:
%Exchange with lead : rate=Gamma^2/E, time=4ns * E/gamma=4ns*500=2000ns. could be enhanced due to electric field excitations
%Exchange with other spin
%\begin{figure}[tr]
\begin{figure}[t]
          \includegraphics[scale=1]{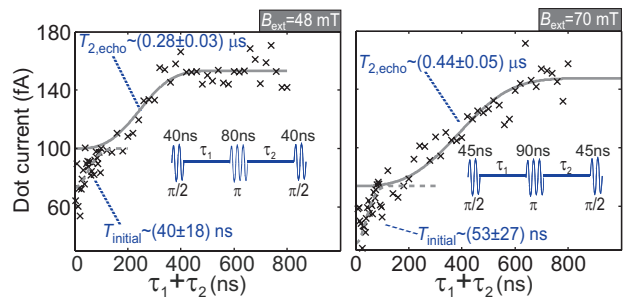}
     \caption{(color online) a) Spin-echo signal at $B_\mathrm{ext}=48$ mT ($f_\mathrm{ac}=280$ MHz) and 70 mT ($f_\mathrm{ac}=380$ MHz). Pulse sequence depicted in the insets. Solid and dashed lines are best fits to the data as in Fig. 3a.}
     
\end{figure}

We thank D. Klauser, W. Coish, D. Loss, R. de Sousa, R. Hanson, S. Saikin,  I. Vink and T. Meunier for discussions; K.-J. Tielrooij for help with device fabrication; R. Schouten, A. van der Enden and R. Roeleveld for technical assistance and L. Kouwenhoven for mentorship and support. We acknowledge financial support from the Dutch Organization for Fundamental Research on
Matter (FOM) and the Netherlands Organization for Scientific Research (NWO).


\begin{thebibliography}{10}
\bibitem{remark_blockade}
Normally, all three triplet states can block current flow. However, due to the
  difference of nuclear fields in the two dots, one of the triplets $T_0$ is
  admixed with the singlet $S$, which lifts the blockade \cite{koppens05}.
  Therefore, only $\ket{\uparrow\uparrow}$ or $\ket{\downarrow\downarrow}$ will
  block the current flow.
\bibitem{vectorsource}
In order to generate sequences of RF bursts with controllable phase, we use a
  vector source (Rohde\&Schwarz SM300, 9 kHz to 3 GHz) with RF modulator,
  controlled by two channels of a Tektronix arbitrary waveform generator (model
  AWG520).
\bibitem{onespin}
For all experiments discussed here, one electron is on-resonance for most of
  the nuclear configurations. This is because on average the nuclear field
  difference between the two dots is larger than the width of the
  power-broadened resonance. Therefore, the driven transitions are mostly
  single spin transitions, although we can not controllably address the spins
  separately. Furthermore, the tunnel coupling between the two dots is small
  enough ($\sim0.05\,\mu$eV) to guarantee (together with the dot alignment)
  negligible Heisenberg coupling $J$ between the spins compared to the
  hyperfine interaction with $\sim10^6$ nuclear spins in the host material.
\end{thebibliography}
\end{document}